\documentclass[usenatbib]{mn2e}

\usepackage{graphics}
\usepackage{epsfig}
\usepackage{natbib}

\begin{document}

\title[Galactic Conformity]{ A Re-examination of Galactic Conformity and a Comparison with Semi-analytic
Models of Galaxy Formation}

\author[G. Kauffmann et al.]{Guinevere Kauffmann$^{1}$\thanks{E-mail: gamk@mpa-garching.mpg.de}, Cheng Li$^{2}$,
Wei Zhang$^{3}$, Simone Weinmann$^{4}$\\
$^{1}$Max-Planck Institut f\"{u}r Astrophysik, 85741 Garching, Germany\\
$^{2}$Max-Planck-Institut Partner Group, Shanghai Astronomical Observatory, China\\
$^{3}$National Astronomical Observatories, Chinese Academy of Sciences, Beijing 100012, China\\
$^{4}$Leiden Observatory, P.O.Box 9513, 2300 RA Leiden, The Netherlands}

\maketitle

\begin{abstract} 
The observed  correlation between star-formation in central galaxies and in their neighbours                         
(a phenomenon  dubbed ``galactic conformity'') is in need of a convincing
physical explanation.     
To gain further insight, we use a volume-limited sample of galaxies with
redshifts less than 0.03 drawn from the SDSS Data Release 7 
to investigate the scale dependence of the effect and how it changes as a function
of the mass of the central galaxy.
Conformity extends over a central galaxy stellar mass range spanning two orders of
magnitude.  The scale dependence and the precise nature of the effect
depend on the mass of the central.  In  central galaxies with masses less than $10^{10} M_{\odot}$,
conformity extends out to scales
in excess of 4 Mpc, well beyond the virial radii of their dark matter halos.
For low mass central galaxies, conformity with neighbours on very large scales is only seen when they  have
low star formation rate or gas content.   
In contrast, at high stellar masses,
conformity with neighbours applies in the gas-rich regime and 
is clearly confined to scales comparable to the
virial radius of the dark matter halo of the central galaxy.
Our analysis of a mock catalogue from the Guo et al (2011) semi-analytic models
shows that conformity-like effects arise 
because gas-poor satellite galaxies are sometimes misclassified as centrals.    
However, the effects in the models are much weaker than observed. Mis-classification only influences the low-end
{\em tail} of the $SFR/M_*$  distribution of  neighbouring galaxies
at large distances from the primary.  The median and the upper percentiles
of the $SFR/M_*$ distribution remain almost unchanged, which is in contradiction with the data. 
We speculate that the conformity between
low-mass, gas-poor central galaxies and their distant neighbours may be a signature  of
``pre-heating'' of the intergalactic gas at an earlier epoch.
The smaller-scale conformity  between high-mass, gas-rich central galaxies and their
close neighbours may be a signature of ongoing gas accretion onto central galaxies
in a minority of massive dark matter halos.

\end{abstract}


\section {Introduction}

In a system of orbiting galaxies, the largest and most massive galaxy is often
referred to as the ``primary'', and the others are called satellites.  
The relationship between satellite galaxies and their primaries is one of the
key tests of hierarchical galaxy formation models. In such models, galaxies form
as cools and condenses at the centers of dark matter halos. As time progresses,
dark matter halos merge and this leads to the formation of systems of galaxies
orbiting within a common potential well.
With the advent of large, wide-field imaging and spectroscopic surveys, there have been
numerous studies of the properties satellite galaxies (e.g. McKay et al 2002;
Prada et al 2003; Sales \& Lambas 2004; Berlind et al 2005; Yang et al 2005;
Weinmann et al 2006; Yang et al 2008; Norberg et al 2008;  More et al 2009)

The behaviour of the average line-of-sight velocity dispersion of satellites  as
a function of distance from the primary galaxy, as well as the number density
distribution of satellites  as a function of projected radius, probes the
density distribution of dark matter in galactic halos.  In general, the velocity
dispersion distributions and  density profiles satellites galaxies are in
reasonably good agreement with the predictions of the $\Lambda$CDM model (Prada
et al 2003; Guo et al 2011). In recent work, Wang \& White  (2012) and 
Sales et al (2012) have shown that the
luminosity and stellar mass functions of satellite galaxies predicted by
semi-analytic models of galaxy formation embedded within high resolution
cosmological N-body simulations agree well with results derived from the Sloan
Digital Sky Survey. 

Our understanding of the observed colours, star formation rates and gas content
of satellites is still not complete. These properties are extremely sensitive to
the physical processes that regulate how gas is supplied to these systems. In
addition, both hydrodynamical and gravitational forces act to remove gas from
satellites. Galaxy formation models developed in the 1990's (e.g. Kauffmann et
al 1993, Cole et al 1994) generally assumed that the diffuse gas halos
surrounding galaxies were stripped instantaneously as soon as they became
satellites.  There is then no further supply of new gas to satellites,
and they redden with respect to the primary galaxy as 
their internal reservoir of cold gas is consumed.  Subsequent work found that these models
predicted satellites that were too red to be consistent with observations
(Weinmann et al 2006; Baldry et al 2006). Current models attempt to model
tidal and ram-pressure stripping of the diffuse gas in a more realistic way
(Font et al 2008; Weinmann et al 2010; Guo et al 2011) and they produce
satellite colour distributions in better agreement with the data.

We note that gas removal processes such as tidal interactions and ram-pressure stripping
operate predominantly on  smaller satellite galaxies, leaving the
interstellar medium of the primary galaxy unperturbed. Only in the case of a
very close encounter, would star formation in the  primary galaxy respond to the
presence of a satellite. The discovery that the properties of satellite galaxies
are strongly correlated with those of their central galaxy, a phenomenon that
has been called ``galactic conformity'' (Weinmann et al 2006), thus remains something
of an enigma.

Yang et al (2006) suggest that a halo that  assembled a significant
fraction of its mass at an early epoch, may have accreted all its satellites at
higher redshifts than a similar halo that assembled late.  They proposed that
galactic conformity may thus be a straightforward manifestation of hierarchical
structure formation and thus ought to be detectable in galaxy catalogues
generated from N-body simulations + semi-analytic models.  
Wang \& White (2012)  analyzed  mock galaxy catalogues generated from the
semi-analytic models of Guo et al (2011). These authors  find a conformity
effect in the models and show that it arises because red central galaxies
inhabit more massive dark matter halos than blue galaxies of the same stellar
mass.  

Other authors have argued that more exotic hydrodynamical effects may be at
play.  Ann, Park \& Choi(2008) argue that feedback processes
that operated as the central galaxy formed
affected the ability of surrounding galaxies to form stars. 
Kauffmann, Li \& Heckman (2010)  suggest that conformity is related to gas accretion.
They propose that gas-rich satellites trace an underlying reservoir of ionized
gas that is extended over large spatial scales, and that this reservoir fuels
star formation in both the satellites and in the primary.

In an attempt to determine which viewpoint is correct, we have undertaken a
re-analysis of galactic conformity using both Sloan Digital Sky Survey data and
the  publically available galaxy catalogues of Guo et al (2011). We have made
the following changes to the analysis procedures that were employed in previous
papers: 

\begin{enumerate} 
\item We restrict the analysis to a volume-limited sample of galaxies
from the spectroscopic catalogue with $\log M_* > 9.25$
and with redshifts in the range $0.017<z<0.03$. Previous
analyses of satellites have made use of much fainter galaxies identified
in the SDSS imaging data without spectroscopic redshifts. 
It has thus been necessary to account for
contamination from galaxies physically unrelated to the primary by means of
statistical background subtraction. Our restriction to a sample of very nearby galaxies with
spectroscopic redshifts decreases the number of primary galaxies we are able to
analyze by large factor, but it produces a much "purer" set of
satellites.   
The quantitative comparison with  semi-analytic models is also greatly simplified.  

\item Because we are not limited by background subtraction errors, we are able to analyze
conformity effects out to projected radii of 4-5 Mpc. The analysis of Kauffmann, Li \& Heckman 
(2010) showed that correlations between satellite and central  properties were
still clearly present at projected separations of $\sim$ 1 Mpc. The analysis of
Wang \& White (2012), was restricted to satellites within a
projected distance of 300 kpc from the primary galaxy, so did not address
the question of whether conformity persists out to large physical scales.  
In addition, we are now able to analyze the {\em full distribution function of specific
star formation rates} in the population of neighbouring galaxies. It is very difficult to
do this accurately using photometric data.   

\item  Previous work has
focused only on average colours, star formation rates and inferred gas fractions
of  satellite galaxies. In this analysis, we look at how {\em relations} between
star formation rate,  stellar mass and structural parameters change for
galaxies located in the vicinity of red and blue primaries.  

\end {enumerate}

In agreement with previous results, we find conformity between the properties
of central galaxies and their neighbours over a central galaxy stellar mass range spanning two orders of
magnitude. However, our new analysis shows that the scale dependence of the effect
depends on the mass of the central.  Conformity effects extend to scales
in excess of 4 Mpc around low mass central galaxies and the strongest effects are seen
at large separations ($>1$ Mpc) from the primary.  In contrast, for high mass
central galaxies,  conformity is clearly confined to scales less than 1-2 Mpc
(i.e. within the scale of the dark matter halo).  
Conformity is only seen for low mass central galaxies when they  have
lower-than-average star formation rate or gas content. 
Conformity  applies when high mass central  galaxies are gas-rich and strongly star-forming. 
The observational results 
are not well-reproduced by the current Guo et al
(2011) semi-analytic models.

Our paper is organized as follows. In section 2, we describe the data used in this analysis.
In section 3, we present results from the SDSS DR7 spectroscopic sample. These results
are compared with the Guo et al. models in section 4. In sections 5 and 6, we summarize and
discuss possible implications of our findings.  
Throughout this paper, we assume a spatially flat concordance cosmology with
$\Omega_m=0.3$, $\Omega_{\Lambda}=0.7$, and $H_0=70$ km s$^{-1}$Mpc$^{-1}$.

\section {Analysis Tools}
\subsection{Data}
We begin with the parent galaxy sample constructed from
the New York University Value Added Catalogue (NYU-
VAGC) sample dr72 (Blanton et al. 2005), which consists of
about half a million galaxies with $r < 17.6$, $-24 < M_r < -16$
and redshifts in the range $0.01 < z < 0.5$. Here, $r$ is the
$r$-band Petrosian apparent magnitude, corrected for Galactic
extinction, and $M_r$ is the $r$-band Petrosian absolute
magnitude, corrected for evolution and K-corrected to its
value at $z = 0.1$.

From the parent sample, we select a volume-limited sample of galaxies
with $\log M_* > 9.25$ and redshifts in the range $0.017 < z < 0.03$.
The lower limit in redshift ensures that we do not consider galaxies where
the peculiar velocity would affect the conversion from redshift into distance
by a significant factor. The upper limit in redshift ensures that we are
able to detect all galaxies down to a limiting stellar mass of $2 \times 10^9 M_{\odot}$,
irrespective of the intrinsic colour of the system. These cuts result in a sample of
11,673 galaxies.  

We define a galaxy with mass $M_*$ to be a central galaxy if there is no other
galaxy with with stellar mass greater than $M_*/2$ within a projected radius
of 500 kpc and with velocity difference less than 500 km s$^{-1}$.
There are  7712 galaxies in our catalogue 
with stellar masses greater than
$5 \times 10^9 M_{\odot}$ to which we can apply this criterion;  
4636 (i.e. 60\%) are classified
as central galaxies. As we will show, this is in reasonably good agreement    
with the predictions of the semi-analytic models of Guo et al (2011).

In this paper, we will make use of two different measures of gas content/star formation
activity.  
\begin {enumerate}
\item The "pseudo" HI gas mass fraction estimates of Li et al (2012)  utilize a  combination of four
galaxy parameters:
\begin{eqnarray}\label{eqn:hiplane}
  \log(M_{\mbox{H{\sc i}}}/M_\ast) & = & -0.325\log\mu_\ast-0.237(NUV-r) \nonumber \\
                     &   & -0.354\log M_\ast-0.513\Delta_{g-i}+6.504,
\end{eqnarray}
where $M_{*}$ is the stellar mass;
$\mu_\ast$ is the surface stellar mass density
given by $\log\mu_\ast=\log M_\ast-\log(2\pi R_{50}^2)$
($R_{50}$ is the radius enclosing half
the total $z$-band Petrosian flux and is in units of kpc).  $NUV-r$ is the global
near-ultraviolet (NUV) to $r$-band colour. The $NUV$ magnitude is
provided by the GALEX pipeline and the $NUV-r$ colour is corrected for Galactic
extinction following Wyder et al (2007) with $A_{NUV-r} = 1.9807A_r$ , where $A_r$
is the extinction in the $r$-band derived from the dust maps of
Schlegel, Finkbeiner \& Davis (1998).
$\Delta_{g-i}$ is the colour gradient defined as the difference
in $g-i$ colour between the outer and inner regions of the galaxy. The inner
region is defined to be the region within $R_{50}$ and the outer region
is the region between $R_{50}$ and $R_{90}$.
As discussed in Li et al (2012), the estimator has been calibrated using samples of nearby galaxies ($0.025<z<0.05$)
with H{\sc i} line detections from the GALEX Arecibo SDSS Survey (Catinella et al 2010).
\item The specific star formation rate (SFR/$M_*$) evaluated within the SDSS fibre aperture is estimated
using the methodology described in Brinchmann et al (2004). These estimates are publically available for
all galaxies in the DR7 galaxy sample at http://www.mpa-garching.mpg.de/SDSS/DR7/.
\end {enumerate}

We note  that the main difference between the two measures is that the first is mainly sensitive
to the age of stars in the outer regions of the galaxy, while the second is sensitive to 
the amount of ongoing star formation in the core of the galaxy (At the median redshift
of the galaxies in our sample, the 3 arcsecond diameter SDSS fibre subtends a physical
scale of only 1.37 kpc).

\subsection {Models}
In this paper we compare our observational results to predictions
from the galaxy formation models of Guo et al (2011; hereafter G11) 
This model was  created by implementing  prescriptions
for baryonic astrophysics on merger trees that follow
the evolution of the halo/subhalo population in the Millennium-II
(MSII; Boylan-Kolchin et al. 2009 ) Simulation, a cubic region
137 Mpc on a side containing 2160$^3$ particles with mass  $9.45 \times 10^6 M_{\odot}$ .
The G11 model is the most recent semi-analytic model
from the Munich group, in which the treatments of many of the physical processes have been
significantly updated. G11 demonstrated that their model provided
good fits not only to the luminosity and stellar
mass functions of galaxies derived from SDSS data, but also
to recent determinations of the abundance of  satellite
galaxies around the Milky Way and the clustering properties
of galaxies as a function of stellar mass. 

In this paper, we work with 13,830 galaxies with stellar masses greater than
$2 \times 10^9 M_{\odot}$ from the $z=0$ output of the simulation.
We identify central galaxies 
in the simulation box in the same way as in the observations, but in the simulation we know which galaxies are
true central and satellite systems, so this allows us to evaluate the efficacy of our procedure.
In the left-hand panel of Figure 1, the thick dashed line
shows the fraction of satellite galaxies F(sat) as a function of stellar mass
for the simulated galaxies; it  
decreases from 0.55 at $M_* = 3 \times 10^9 M_{\odot}$ to around 0.25 at
$M_*= 3 \times 10^{11} M_{\odot}$. The thin lines show F(sat)
once the isolation cuts have been applied.   
Red, black, green, blue and cyan
lines are for galaxies with increasing cold gas fractions.  
At the low mass end, the isolation cuts decrease F(sat) by
a factor of between 2 and  5, depending on gas fraction.  
At the high mass end, the cuts have much smaller effect. 
This is because the relation between 
central galaxy and dark matter halo mass is rather flat for high mass halos
and as a result, the true central galaxy is not always the most massive galaxy
in its immediate environment. Nevertheless, we see that the predicted contamination
from satellites is always below 30\% for galaxies of all stellar masses and gas fractions.

\begin{figure}
\includegraphics[width=58mm]{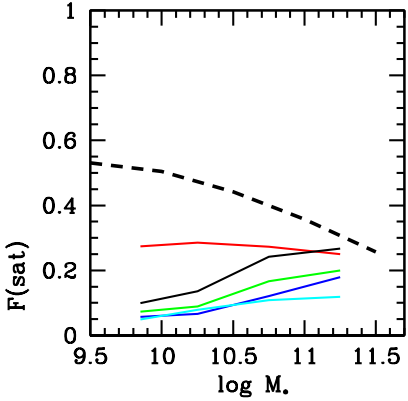}
\caption{ The thick dashed line shows the fraction of satellite galaxies as 
a function of stellar mass for all galaxies in the $z=0$ Millennium II
simulation output.  The coloured lines show the fraction of satellites
once we apply the same isolation criterion used  to construct our SDSS
central galaxy sample (see text for details). Red, black, green, blue and cyan curves
are for simulated galaxies with cold gas mass fractions in the 0-25th percentile,
25-50th percentile, 50-75th percentile, $>$ 75th percentile and $>$ 90th percentile
ranges of the full distribution.    
\label{sample}}
\end{figure}

\section {Results from SDSS}

\subsection {Dependence of conformity on the stellar mass of the primary, separation
of the neighbour, and star formation activity tracer}

In this section, we carry out a systematic exploration of how
conformity between central galaxies and the surrounding population
of neighbours depends on, a) the stellar mass of the central, b) the
physical separation between the neighbour  and the central, c) the indicators
used to trace star formation and cold gas content in both the centrals and in their
neighbours.  

Figures 1-4 in this section focus on how the specific star formation
rates and gas content of the neighbouring galaxy  population vary as a function of
projected radius from the central. Figures 5-6 then explore how the {\em relations}
between specific star formation rate/gas content and galaxy mass/structural
parameters for neighbouring  galaxies change according to the properties of the central object.

We begin with the subset of central galaxies with stellar masses in the range
$10.0 < \log M_* < 10.5$. We divide these galaxies into quartiles using four different measures:
\begin {enumerate}
\item The "pseudo" HI mass fraction estimate given in equation 1. Hereafter, we will
denote this quantity as GS($M_*$).   
\item  The  "HI deficiency parameter" defined in Li et al. (2012), which is 
the deviation in $\log(M_{HI}/M_*)$ from the value
predicted from the mean relation between $\log(M_{HI}/M_*)$ and the combination of
galaxy mass $M_*$ and stellar surface mass density $\mu_*$ . Li et al (2012) showed
that the amplitude of the correlation function on scales of 1-2 Mpc varied much
more strongly as a function of the HI deficiency parameter than as a function
of the "pseudo" HI mass fraction. In subsequent work, Zhang et al (2012) found that
depletion of gas in galaxies in  groups and clusters depended more strongly on the stellar
surface density of the galaxy than on the  mass of the galaxy. 
It is thus instructive to fix {\em both stellar mass and galaxy size} when
we investigate how galaxies are affected by their environment.
\item The specific star formation rate SFR/$M_*$ evaluated within the 3 arcsecond diameter
fibre aperture. As we have discussed,  
this quantity should be regarded as a measure of
recent star formation activity in the central regions of the galaxy.
\item The total specific star formation rate. We note that our estimate
of SFR(total) is dominated by the aperture correction to SFR(fibre). In Brinchmann et al (2004),
which was based on SDSS DR4 data,  the
aperture corrections were done in an empirical manner. The procedure was later found to over-estimate
the SFR of red  galaxies (Salim et al 2007). In the DR7 release, the aperture corrections are
computed by fitting the SDSS 5-band photometry of the outer galaxy to a library of 
spectral energy distributions  generated using
population synthesis models.  
\end {enumerate}

In Figure 2,  results for galaxies in  four quartiles of            
HI gas mass fraction, HI deficiency parameter, fibre and total specific SFR/$M_*$
are shown as red, black, green and blue curves, respectively. The cyan curves show results for
galaxies in the upper 90th percentiles of these quantities, i.e. cyan curves are for
very unusually gas-rich or strongly star-forming central galaxies.
The solid curves show the median fibre-based specific star formation rate of neighbouring
galaxies as a function of projected distance (in Mpc) from the central object. 
The lower and upper set of dotted lines
show the 25th and 75th percentiles this quantity.                               

Fow all central galaxy bins, the specific star formation rates of neighbours as a function
of projected radius first exhibit a drop, and then flatten out at radii larger than 500 kpc.
This scale is artificially {\em imposed} by our definition that  a central galaxy
has to be a factor of two  more massive than any neighbour within a projected radius of 500 kpc. 
Low mass galaxies have higher specific star formation rates than high mass galaxies,
so by eliminating any galaxies with massive companions within $R_{proj}= 500$ kpc, the specific star
formation rates are forced to change in a discontinuous way at this radius.   

It is apparent that for central galaxies with stellar masses $\sim 10^{10} M_{\odot}$,
conformity between  central galaxies and neighbours is strong if the centrals have
gas mass fractions/specific star formation rates less than the median value. It is
absent for central  galaxies with gas fractions and SFR/$M_*$ values
higher than the median.  It is also apparent that conformity  extends
out to very large projected separations. In all four panels, the black curve does not converge to meet the
cyan, blue and green curves until a projected radius of  $R=3$ Mpc. The red curve,
which shows results for the 25\% most gas-poor/weakly star-forming galaxies, remains depressed
out beyond $R=4$ Mpc. These are scales well beyond the virial
radius of a $10^{10} M_{\odot}$ galaxy! These conclusions are independent of the choice 
of star formation activity indicator.
Finally, conformity is most pronounced when the central
galaxies are ordered by global HI deficiency and total specific star formation rate,
rather than central $SFR/M_*$. 

Figure 3 is the same as Figure 2, except that results are shown for  central galaxies with stellar
masses in the range $10^{11} - 3 \times10^{11} M_{\odot}$. There are many fewer galaxies
in this stellar mass bin, so the errorbars are larger, but it is nevertheless clear that the 
results are strikingly different.
First, conformity effects are only apparent for galaxies in the upper quartile
of HI gas mass fraction, HI deficiency and total specific star formation rate. Second, unlike Figure 2,
no conformity is  seen if the central galaxies are ordered by central specific
star formation rate. Third, conformity is  largest at small separations
and disappear at projected radii beyond $\sim 2-3$ Mpc.

\begin{figure}
\includegraphics[width=95mm]{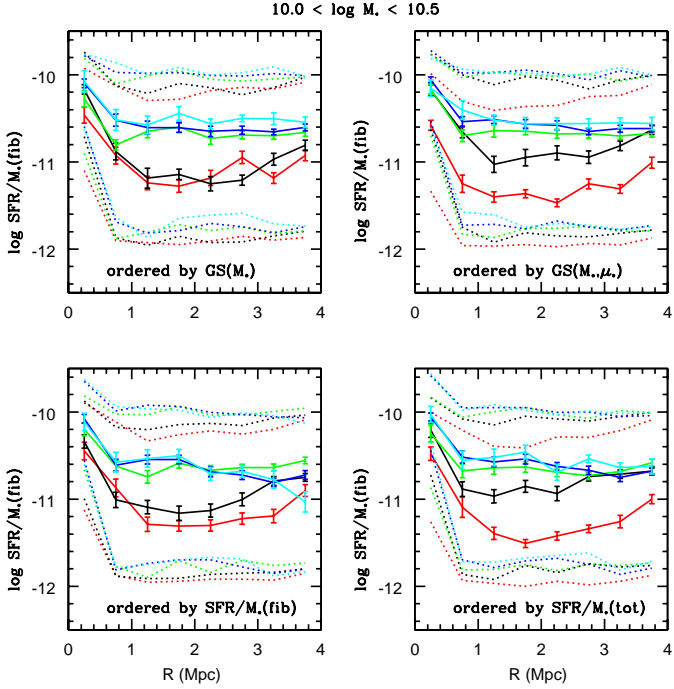}
\caption{ The specific star formation rate (measured within the SDSS fibre aperture)
of neighbouring galaxies is plotted as a function of projected distance from
the central galaxies. Results are shown for central galaxies in the stellar
mass range $10 < \log M_* < 10.5 M_{\odot}$. In each of the four panels, the central
galaxies have been ordered by a different quantity: a) "pseudo" HI mass fraction (top left),
b) HI deficiency parameter (top right), c) fibre specific star formation rate (bottom left),
d) total specific star formation rate (bottom right).  
Red, black, green, blue and cyan curves indicate results for central  galaxies that fall into the   
 0-25th percentile,
25-50th percentile, 50-75th percentile, $>$ 75th percentile and $>$ 90th percentile
ranges of distribution of these four quantities. Solid curves indicate the median of
the SFR/$M_*$ distribution for neighbouring galaxies at given radius,  while upper and
lower dotted curves indicate the 25th and 75th percentiles of the  SFR/$M_*$ distribution.
Errorbars on the median have been computed via  boot-strap resampling.    
\label{sample}}
\end{figure}

\begin{figure}
\includegraphics[width=95mm]{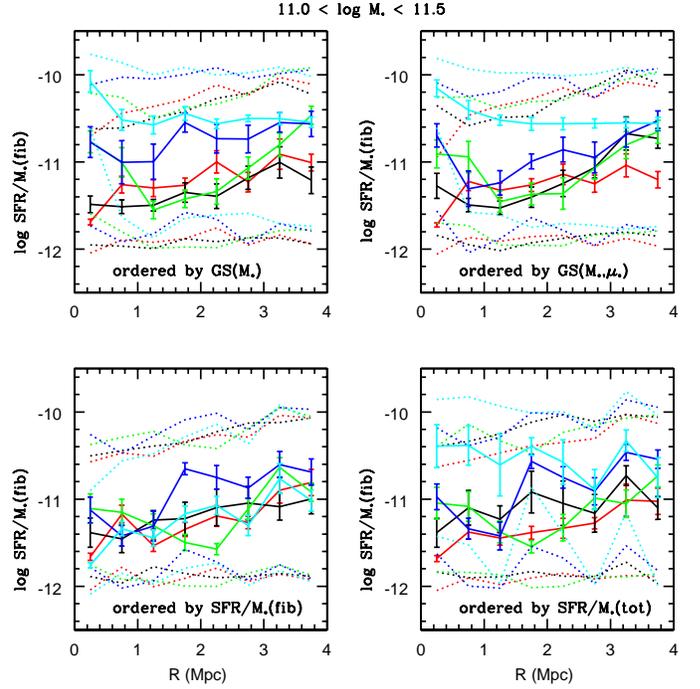}
\caption{ As in Figure 2, except for central galaxies in the stellar                                 
mass range $11 < \log M_* < 11.5 M_{\odot}$. 
\label{sample}}
\end{figure}

In Figure 4, we investigate trends with stellar mass in more detail by showing results in four different
central galaxy stellar mass bins spanning the range $M_*= 5 \times 10^9 M_{\odot}$ to $3 \times 10^{11} M_{\odot}$. 
For simplicity, we only show the case where the central galaxies are ordered by HI deficiency. 
At low stellar masses, conformity is  strongest on large scales
and applies only in the gas-poor regime. 
At high stellar masses, conformity is strongest on small scales and applies only in the gas-rich
regime. The cross-over between the two regimes occurs for central
galaxies with stellar masses $\sim 3 \times 10^{10} M_{\odot}$. In the $10.5 < \log M_* < 11 M_{\odot}$ bin,
conformity is seen both on small scales for gas-rich central galaxies, and on large scales for
gas-poor central galaxies. 

\begin{figure}
\includegraphics[width=95mm]{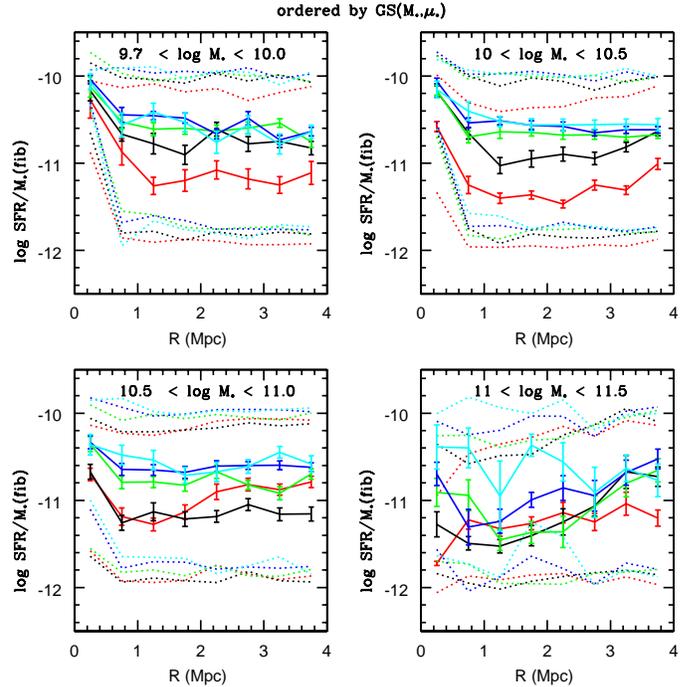}
\caption{ As in Figure 2, except results are shown for central galaxies in four different stellar mass ranges.
For simplicity, we only show the case where central galaxies are ordered by HI deficiency.
\label{sample}}
\end{figure}

So far, we have only investigated the sensitivity of conformity  to 
the indicator used to partition central galaxies into gas-rich/strongly-star-forming and
gas-poor/weakly-star-forming systems. We concluded that conformity is  strongest  
when the central galaxies are sorted according to global quantities 
(total gas content and total specific star formation rate). 
What about the neighbours? Figure 5 is the same as Figure 4, except
that we plot the pseudo gas fractions of the neighbours instead of their central specific star
formation rates. 
The amplitude of conformity effect is now much smaller. In the 
$10<\log M_* < 10.5 M_{\odot}$ bin, the difference in the median $SFR/M_*$(fibre) for neighbours 
around the most gas-rich centrals compared the most gas-poor centrals is nearly a factor of 10.  
In contrast, the difference in the  gas fraction 
is less than a factor of 2.  This decrease  in the amplitude 
holds in all four stellar mass bins. 

What we conclude, therefore,  is that an excess or deficiency
in the gas content
of central galaxies is most intrinsically correlated  with the timescale over which
their neighbours have been building their {\em central} stellar populations.

\begin{figure}
\includegraphics[width=95mm]{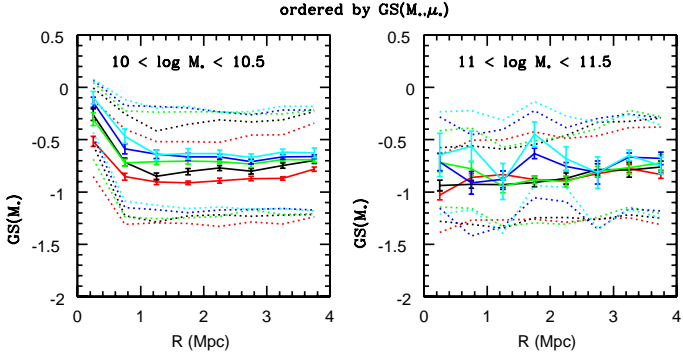}
\caption{ As in Figure 4, except results are shown for central galaxies in two different stellar mass ranges
and we plot the pseudo gas fractions of their neighbours, rather than their fibre specific star
formation rates.
\label{sample}}
\end{figure}

\subsection {Correlations between different galaxy properties in neighbouring galaxies}

In the previous section, we examined how the median specific star formation rates and gas fractions of 
neighbouring galaxies at a given projected radius correlated with the properties of the
central object. In this section, we study changes in the {\em relations} between
different satellite  properties such
as stellar mass, stellar surface density, concentration index, specific star formation rate
and gas mass fraction. 

As discussed in the previous subsection, conformity is strongest on large scales
for low mass centrals, and on small scales for high mass centrals.
Here we analyze central galaxies in two different stellar mass ranges:
$9.7 < \log M_* < 10.3 M_{\odot}$ and $10.7 < \log M_* < 11.5 M_{\odot}$. For the lower mass bin, 
we pick all neighbours  with projected radii between 1 and 3 Mpc and
$\Delta cz < 500$ km/s and we plot relations
between different properties in Figure 6. For central galaxies in the higher mass bin,
we pick satellites with projected radii less than 0.6 Mpc and plot
the corresponding relations in Figure 7. 
In both plots, red, black, green and blue curves denote median relations for
neighbours around central galaxies divided into four quartiles 
in HI deficiency parameter.

\begin{figure}
\includegraphics[width=95mm]{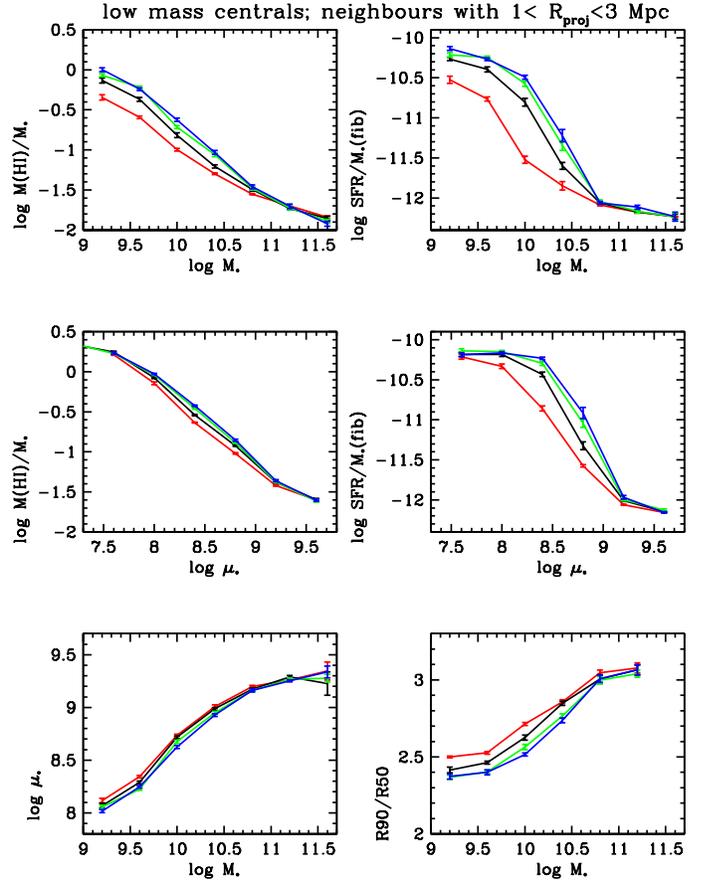}
\caption{ The relations between pseudo gas mass fraction and stellar mass (top left),
fibre specific star formation rate and stellar mass (top right), pseudo gas mass
fraction and stellar surface mass density (middle left), fibre specific star formation rate
and stellar surface mass density (middle right), stellar surface mass density and stellar
mass (bottom left) and concentration and stellar mass (bottom right), for distant neighbours of
central galaxies with stellar masses in the range $9.7 < \log M_* < 10.3$.
Distant neighbours are defined to be at projected radii between 1 and 3 Mpc from the central
galaxy and to have velocity difference $\Delta (cz) <$  500 km/s.  
Red, black, green, blue and cyan curves
are for central galaxies with HI deficiency parameters  in the 0-25th percentile,
25-50th percentile, 50-75th percentile and $>$ 75th percentile 
ranges of the full distribution.
\label{sample}}
\end{figure}

\begin{figure}
\includegraphics[width=95mm]{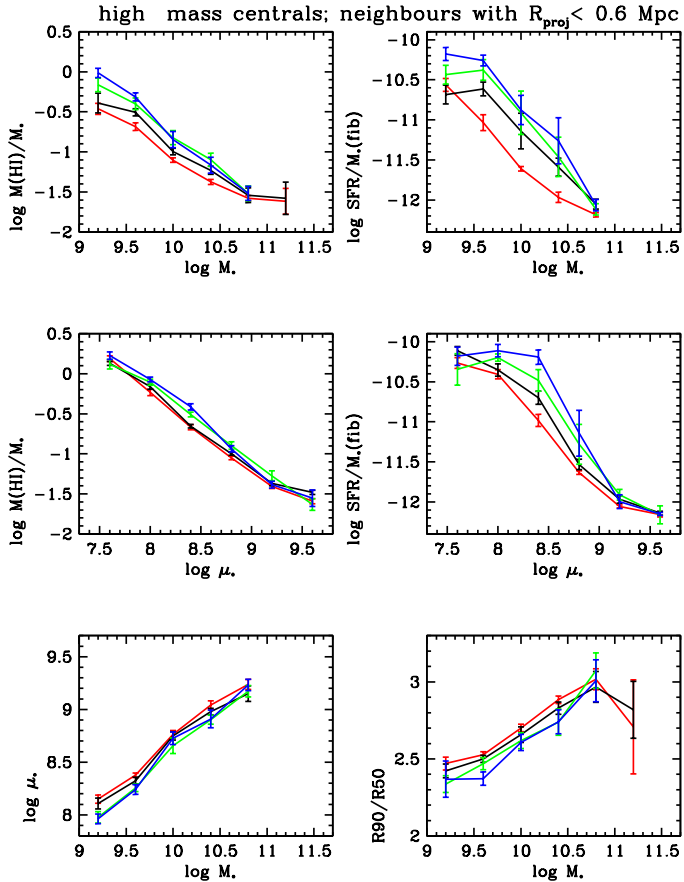}
\caption{ As in Figure 6, except for near neighbours of
central galaxies with stellar masses in the range $10.7 < \log M_* < 11.5$.
Near neighbours are defined to be at projected radii less than 600 kpc  from the central
galaxy and to have velocity difference $\Delta (cz) <$  500 km/s.  
\label{sample}}
\end{figure}

Echoing results presented in the previous section, we see that conformity 
effects only apply in the low-gas fraction regime for low mass central galaxies (i.e.
the blue and green curves are almost indistinguishable in Figure 6). For high mass central
galaxies, conformity  effects extend into the high gas mass fraction regime.
In both Figures 6 and 7, we see that the largest changes occur in the relations between
specific star formation rate measured in the fibre and galaxy mass and stellar surface density. 
Relations between structural parameters such as concentration and stellar surface density
and stellar mass change more weakly. The largest changes in specific star formation rate
occur for neighbours with stellar masses below a few $\times 10^{10} M_{\odot}$ 
and with stellar surface mass densities of a 
few $\times 10^8 M_{\odot}$ kpc$^{-2}$. In this regime, the median value of
SFR/$M_*$ changes from values around $1/15.$ Gyr$^{-1}$ for neighbours around gas-rich central galaxies,
to values around $1/100$ Gyr$^{-1}$ for neighbours around gas-poor central galaxies. In other words,
if such a neighbour is found around a gas-rich central, it is  building up its central stellar mass
over timescales comparable to the Hubble time. If it is found around a gas-poor central,
the growth time of the central region of the galaxy is an order-of-magnitude longer.
It is also interesting  that the galaxy parameter regime where conformity effects
are strongest is the same as the parameter regime of host galaxies of actively
accreting present-day black holes (Heckman et al 2004).

Finally, Figure 8 is a simple summary of what we believe to be the main result of this section. 
We compare the systematic changes in the relations between fibre specific star formation rate
and stellar surface mass density for distant
neighbours around low mass centrals (left) with those for near neighbours around high
mass centrals (right). Note that we have added 
central galaxies in the upper 90th percentile in HI content as a cyan curve in each panel. 
The difference in behaviour in the two panels as a function of the gas content
of the central galaxies is quite striking. 
We will discuss possible interpretations in the final section.

\begin{figure}
\includegraphics[width=95mm]{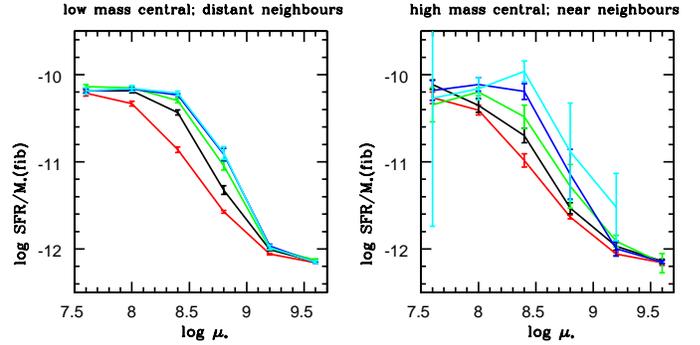}
\caption{ The relations between 
fibre specific star formation rate and stellar surface mass density 
for distant neighbours of
central galaxies with stellar masses in the range $9.7 < \log M_* < 10.3$ (left)
and for close neighbours of
central galaxies with stellar masses in the range $10.7 < \log M_* < 11.3$ (right).
Red, black, green, blue and cyan curves
are for central galaxies with HI deficiency parameters  in the 0-25th percentile,
25-50th percentile, 50-75th percentile, $>$ 75th percentile and $>$ 90th percentile
ranges of HI deficiency parameter.
\label{sample}}
\end{figure}

\section {Results from semi-analytic models}
In this section, we present results from the semi-analytic model galaxy catalogues. 
As discussed in the introduction, Wang \& White (2012) have analyzed conformity
in these models and claim that the observed trends are explained because red galaxies inhabit more massive
halos than blue galaxies of the same mass. We do not disagree with the statement
that the halo masses of red and blue galaxies of the same $M_*$ are different . As seen in Figure 1,
the fraction of central galaxies that are mis-classified even after our isolation criterion 
is applied is significantly higher for gas-poor galaxies -- all of these mis-classified
centrals are satellite systems in massive halos.  However, as we will show in this section,
{\em this effect cannot explain the trends seen in the observations.}     
 
The Guo et al (2011) model does not include detailed modelling of the density profiles of the
stars and gas, so we only work with global specific star formation rates and gas fractions
in this section. 
Figure 9 is analogous to Figure 2: we plot the specific star formation rates of neighbours  as a function 
of projected distance from central galaxies with stellar masses in the range $10 < \log M_* < 10.5 M_{\odot}$.
As in the previous section, we partition the central galaxies into quartiles in gas mass fraction
(left) and global specific star formation rate (right). Solid curves show the median value of
$\log SFR/M_*$ for the satellites as a function of projected radius, while the lower and upper dotted curves
show the 25th and 75th percentiles of the distributions. As in previous plots,
red, black, green, blue and cyan curves colour-code results for central galaxies with increasing
gas fraction and $SFR/M_*$.

\begin{figure}
\includegraphics[width=95mm]{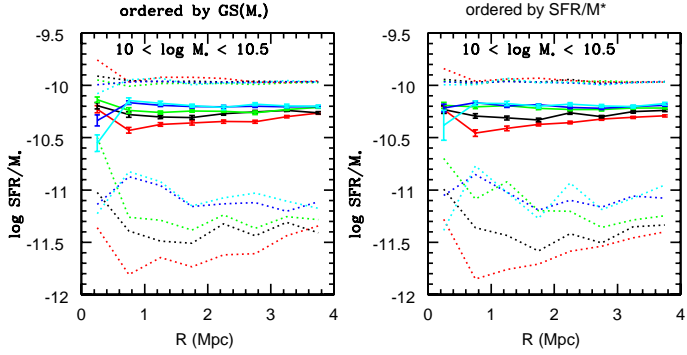}
\caption{ The specific star formation rate 
of neighbouring galaxies is plotted as a function of projected distance from
central galaxies identified in galaxy catalogues generated using the Guo et al (2011)
semi-analytic models. Results are shown for central galaxies in the stellar
mass range $10 < \log M_* < 10.5$. The central
galaxies have been ordered by cold gas mass fraction (left),
and specific star formation rate (right).  
Red, black, green, blue and cyan curves indicate results for central  galaxies that fall into the   
 0-25th percentile,
25-50th percentile, 50-75th percentile, $>$ 75th percentile and $>$ 90th percentile
ranges of distribution of these four quantities. Solid curves indicate the median of
the SFR/$M_*$ distribution for neighbouring galaxies at a given radius, while upper and
lower dotted curves indicate the 25th and 75th percentiles of the  SFR/$M_*$ distribution.
\label{sample}}
\end{figure}

Similar to what is found in the observations, conformity mainly applies
in the low gas fraction/ specific star formation rate  regime. However,  
in the data, the {\em median} specific star formation rate of the neighbours shifts by a factor
of $5-10$ between gas-poor and gas-rich central galaxies and the upper 75th perctiles
of the distribution also show a pronounced effect. In the models,
the median shifts by less than 50\% and the upper 75th percentile shows
no effect whatsoever; the main shift is in the low SFR/$M_*$ tail of the distribution.
This is because conformity effects in the models are caused by the increasing fraction of true satellite galaxies
among objects classified as centrals using our isolation criterion.  However, Figure 1 shows that   
these mis-classifications never come to dominate the statistics, even
in the case of the most gas-poor objects. 

In Figure 10, we show model results for central      
galaxies in the same four bins of stellar mass as in Figure 4. The shifts in median SFR/$M_*$
for neighbouring galaxies  are very weak at all stellar masses. Stronger effects are
seen for the lower 25th percentile of the distribution. 
Conformity also  disappears completely at central galaxy stellar masses
greater than $10^{11} M_{\odot}$, which again disagrees with the data.

\begin{figure}
\includegraphics[width=95mm]{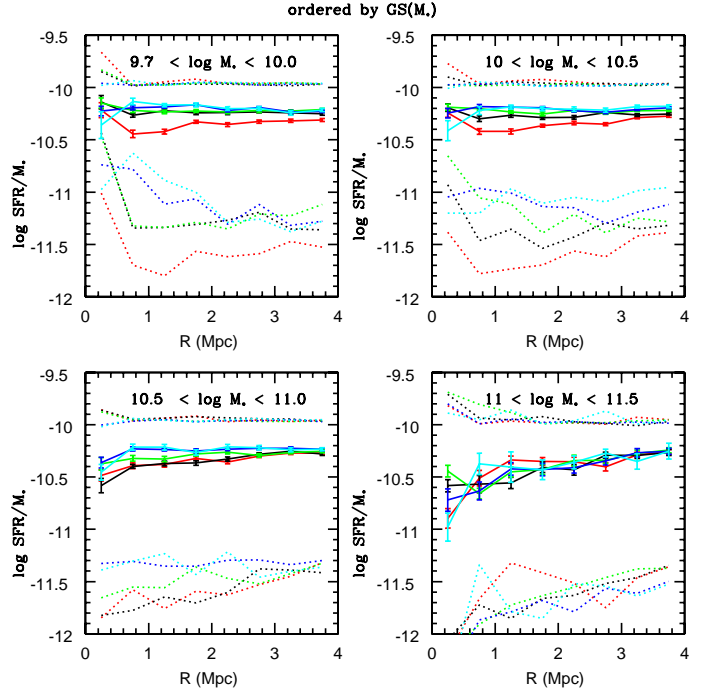}
\caption{ The specific star formation rate 
of neighbouring galaxies is plotted as a function of projected distance from
central galaxies in the model. Results are shown for central galaxies in four stellar mass ranges. 
The central
galaxies have been ordered by cold gas mass fraction. 
\label{sample}}
\end{figure}

Finally, the top two panels of Figure 11 show the relations between
specific star formation rate and stellar mass for neighbours around low mass and high
mass central galaxies. Once again, we 
see no shift in the median relation for neighbours around gas-rich and gas-poor central
galaxies -- only the lower percentiles of the distribution show a significant trend.
These results disagree  with those presented in Figures 6, 7 and 8.

So far we have established that there is very significant disagreement between the data and
the models. Can we now use the simulations to gain insight into what has to be changed
to obtain a better match to the observations? One question we might ask is whether
we need to modify the physical processes regulating gas accretion and star formation
in galaxies that reside at the centers of their dark matter halos, or whether it is
the treatment of processes such as ram-pressure and tidal stripping in infalling
satellites that needs changing, or both.

The middle panels of Figure 11  show the fraction of neighbouring  galaxies 
that are {\em true satellites} in the simulation, i.e. the fraction that do not
reside at the centers of their host dark matter halos. In the middle right panel,
F(sat) is very close to unity, because we have only included galaxies in the 
close vicinity ($R < 600$ kpc) of massive centrals.  The            
conformity between gas-rich massive galaxies and near neighbours seen in 
the right panel of Figure 8 could thus plausibly be recovered by suitable
changes to the recipes for gas-stripping in the simulation.   

In the left panel of Figure 11, we are considering neighbours out to much larger 
projected radii and  F(sat) is around 0.5 on average.
The fraction of true satellites among neighbours does depend on the gas fraction
of the central object, but the effect is quite weak -- F(sat)  shifts
by less than 50\% between the most gas-rich and gas-poor central galaxies.   
Let us assume for argument's sake  that
all satellites are passive and have log SFR/$M_*$ $\sim 10^{-12}$, and that all centrals are
active with  log SFR/$M_*$ $\sim 10^{-10}$; a change in satellite fraction from
40\% to 70\% would only change the average in log SFR/$M_*$  by 0.2 dex, which
is far less than what is seen in Figure 6.

The bottom left panel shows the relation between halo mass and stellar mass for
neighbouring galaxies around low mass centrals that are themselves central objects in their
dark matter halos. In the models, this relation is quite insensitive to  
whether the low mass central is gas-rich or gas-poor. We therefore conclude that
the conformity between  gas-poor central galaxies with low stellar masses and
distant neighbours requires quenching processes to be coordinated in 
galaxies occupying {\em disjoint
and widely-separated dark matter halos}.

\begin{figure}
\includegraphics[width=95mm]{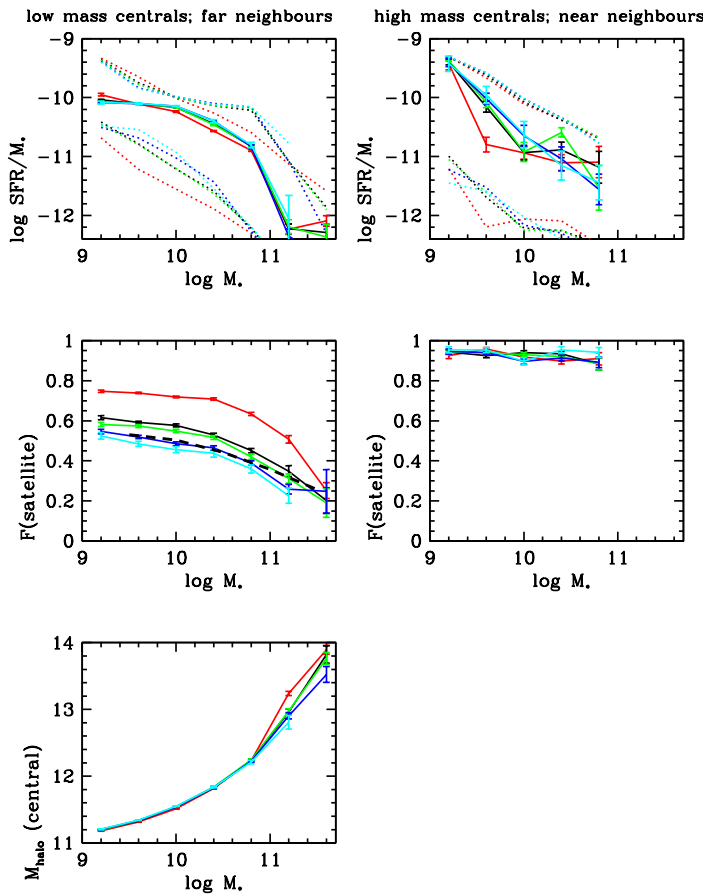}
\caption{ {\bf Top:} For comparison with Figure 8, the relation between specific star formation rate and stellar mass
is plotted for distant neighbours around low mass central galaxies (left) and for
near neighbours around high mass central galaxies (right) in the models.   
Coloured lines indicate different ranges in central galaxy cold gas mass fraction, as before.
{\bf Middle:} The fraction of true satellite galaxies as a function of stellar mass
in the distant  neighbour population around low mass centrals (left);
the same for the near neighbour population around high mass central (right).  
{\bf Bottom left:} Dark matter halo mass is plotted against stellar mass for true central
galaxies in the distant  neighbour population around low mass centrals.  
\label{sample}}
\end{figure}

\section {Summary}

In this paper, we use a volume-limited sample of galaxies drawn from the SDSS
Data Release 7  with  stellar masses  greater than
$ 2 \times 10^9 M_{\odot}$ and redshifts less than 0.03 to perform a detailed analysis
of the ``conformity'' 
between the star formation rates of central galaxies and those 
of neighbouring galaxies pointed out by Weinmann et al (2006).   
We investigate the scale dependence of the effect and how it changes as a function
of the mass of the central. We also explore conformity a variety of different galaxy properties
clarify which ones result in the strongest correlations between central galaxies their
neighbours. Finally, we test whether current semi-analytic models
are able to match the data. 

Our main results are as follows:
\begin {itemize}
\item Conformity between the properties
of central galaxies and their neighbours extends over the full central galaxy stellar 
mass range that we were able to explore ($5\times 10^9 M_{\odot}$ to
$3 \times 10^ {11} M_{\odot}$). 
\item The scale dependence of the effect
depends on the mass of the central.  Conformity extends to scales
in excess of 4 Mpc around low mass central galaxies and the strongest effects are seen
at large separations ($>1$ Mpc) from the primary, well beyond the virial radius
of its dark matter halo.  In contrast, for high mass
galaxies,  conformity is clearly confined to scales less than 1-2 Mpc.
\item For low mass galaxies,
conformity is only seen when the central galaxies have
lower-than-average star formation rate or gas content. For high mass galaxies,
conformity  applies in the high gas fraction/specific star formation rate regime.
\item The strongest conformity effects arise when  gas in central
galaxies is correlated with central specific star formation rates in neighbouring galaxies
with  stellar masses less than a few $\times 10^{10} M_{\odot}$ and stellar
surface densities in the range $10^8 -10^9 M_{\odot}$ kpc$^{-2}$.       
\item Conformity effects in the Guo et al (2011) models are much weaker than the ones that we observe.
The effects in the models occur because a higher fraction of gas-poor galaxies are mis-classified 
as centrals, even after our isolation cut is applied. 
Matching the data for low mass central galaxies requires quenching processes to be coordinated in
galaxies occupying  disjoint
and widely-separated dark matter halos.
\end {itemize}

\section {Discussion}

In this section, we attempt to provide a physical interpretation of our results. We also
speculate on why there is such strong disagreement between the models and the data. Finally, we discuss
possible implications of our results for large-scale structure studies.  
  
\subsection {Low mass galaxies and ``pre-heating''.}
Perhaps the most striking and puzzling new result presented in this paper is that
conformity between low mass central galaxies and their neighbours extends over
scales of many megaparsecs. Conformity  applies to low mass centrals that have gas fractions 
lower than the median value, suggesting that ``quenching'' rather than accretion processes are at work.  

Wang et al (2009) suggested that red, low mass central galaxies are actually satellites that
have been "ejected" from massive halos after undergoing tidal and/or ram-pressure stripping.  
A full explanation of the conformity effect presented in this 
paper would require {\em more than half} of all low mass
galaxies to have passed through a massive halo, which we deem to be unlikely. In addition,
ejection of satellites would likely produce a disjoint population of red, low mass field galaxies
around massive halos. As shown in the left panel of Figure 8, the median SFR/$M_*$ of distant neighbours
changes in a smooth way between the central galaxy bins indicated by green, black and red curves.
This is difficult to understand in the context of ejection.

An alternative possibility is that inter-galactic gas was heated over large spatial scales at some earlier
epoch, preventing it from collapsing into halos, cooling and forming stars.   
In the literature, this has traditionally been referred to as ``pre-heating". Pre-heating was
originally proposed to explain why hydrodynamical simulations failed to
reproduce the cluster temperature-X-ray luminosity relation (e.g. Valageas \& Silk 1999).
Implications of an early epoch of pre-heating on the present-day galaxy population were explored
by Mo \& Mao (2002). These authors postulated  that vigorous energy feedback associated
with star formation and AGN activity in galaxies at redshifts 2-3 would be  responsible for the pre-heating.
In subsequent work, Mo et al. (2005) explored the possibility that shocks produced 
as a consequence of gravitational collapse of large-scale structure could also heat the gas
and prevent it from collapsing into halos. The predictions in this paper have not yet been verified using
hydrodynamical simulations. The fact that conformity is only found for the neighbours of gas-poor
galaxies with {\em low stellar masses} may point to energy sources that are internal rather
than external to galaxies -- supernovae and/or winds from accreting black holes will more
easily reach the IGM if they are generated in low mass halos. 

In the semi-analytic models, pre-heating of the IGM is modelled in a very rough way by 
placing some fraction of the gas in the halo into a so-called "ejected component" (De Lucia,
Kauffmann \& White 2004). This gas is not available for cooling until it is returned to
the dark matter. In the Guo et al (2011) models, gas in the ejected component is returned
to the halo in a few dynamical times for a Milky Way mass galaxy, but on longer timescales
for lower mass galaxies. These models nevertheless overpredict the abundance of
galaxies with stellar masses less than $\sim 10^{10} M_{\odot}$ at redshifts greater
than $\sim 0.6$. Recent work has suggested that the problem of the
overproduction of low mass galaxies at high redshifts is alleviated if 
the gas re-incorporation timescales are longer than assumed by
Guo et al., particularly at higher redshifts (B. Henriques and S. White, private communication).
It is unlikely that such models will do a better job of  matching the observed 
conformity effects, because the ejected component is always returned to the same halo. 
If the ejected compoents of neighbouring galaxies were allowed to {\em mix}, it may be possible
to obtain effects similar to those seen in the data.

\subsection {High mass galaxies and accretion}
Conformity between high mass central galaxies and their neighbours extends over spatial
scales comparable to that of an individual massive dark matter halo.
Conformity is strongest for a minority of the most gas-rich massive galaxies (see right
panel of Figure 8), suggesting that accretion processes may be the underlying cause.  

In Kauffmann, Li \& Heckman (2010), it was argued that blue, star-forming satellites
trace an underlying reservoir of ionized gas that provides fuel for ongoing
star formation in central galaxies. Accretion of both  dark matter and gas from
the surrounding environment  is modelled in detail in the
semi-analytic models, so the question arises as to why  similar conformity effects are not seen
in the high stellar mass bins in Figure 10. 

In the simulations, central galaxies with stellar masses    
of $\sim 10^{11} M_{\odot}$  reside in dark matter halos with masses of few $\times 10^{12} M_{\odot}$
(see bottom left panel of Figure 11).
In the models, gas cooling in halos of this mass occurs from a corona of hot gas
that is assumed to be in hydrostatic equilibrium with the surrounding dark matter halo. Cooling rates 
are regulated by ``radio-mode feedback'', with efficiency that scales with the black hole
mass of the central galaxy multiplied by $V_{vir}^3$, where $V_{vir}$ is the circular velocity of
the dark matter halo (Croton et al 2006). Cooling rates are thus  
mainly determined by the mass of the halo in which the galaxy resides. Because the cooling 
equations {\em assume} that the gas is always in equilibrium with the dark matter, cooling rates
are not sensitive to the dynamical state of the halo or to its satellite population.         

One possibility is that massive galaxies surrounded by blue satellites reside in massive
dark matter halos that have assembled relatively recently. Gas has not yet shocked and reached very
high temperatures and is thus able to cool more efficiently onto the central object.  
The fact that the blue satellite population is most pronounced in the extreme tail of objects with
the highest gas mass fractions, supports the notion that gas fuelling associated with blue
satellites is a transient phase in the lives of these galaxies.
Cooling and star formation may later be shut down by radio jets (see Chen et al 2012).

One might  ask why we do not observe any link between gas-rich central galaxies of low stellar masses and 
blue/star-forming neighbours. This might be expected in a scenario in which star formation in low mass galaxies 
is fuelled by accretion of cold gas from surrounding filaments. 
However, as we have discussed,  the dominant source of gas infall at low redshifts may be in
the form of material previously
ejected by supernovae-driven winds (see also Oppenheimer et al 2010). The extent to which  this gas would
correlate with the present-day population of galaxies is currently not understood.

\subsection {Implications for large-scale structure studies}
We note that a large fraction of current work in cosmology is based on the 
premise that the statistical properties of a galaxy population can be predicted
if one knows just two things: a) the mass distribution of the dark matter halos that host
the galaxies, b) the location of the galaxies in their halos -- in particular whether they
are central galaxies or satellites. This forms the basis of the so-called halo
occupation distribution (HOD) modelling technique. The HOD framework also lies at the root 
of the idea that galaxies trace the underlying distribution in a simple enough way that
they can be used as cosmological probes with high precision.

Most analyses that have tested the halo model have used luminosity or stellar mass-selected
samples as their basis. However, future Baryon Acoustic Oscillation experiments such as
HETDEX and BigBOSS aim to work with high-redshift galaxies selected by their emission line
properties. It is possible that the large-scale conformity effects discussed in this paper
will affect clustering statistics in emission-line-selected galaxy surveys more than in stellar   
mass-selected galaxy surveys. In Figure 12, we show the median as well as the 25th and 75th
pecentiles in the {\em stellar masses} of neighbouring galaxies as a function of
projected radius from the central. As can be seen, there is {\em no conformity} between the gas fraction of
centrals and the stellar masses of neighbours both in the data and in the models. As   
is already  known (Guo et al 2011), the
stellar mass distributions in the data and the models agree very well.

\begin{figure}
\includegraphics[width=95mm]{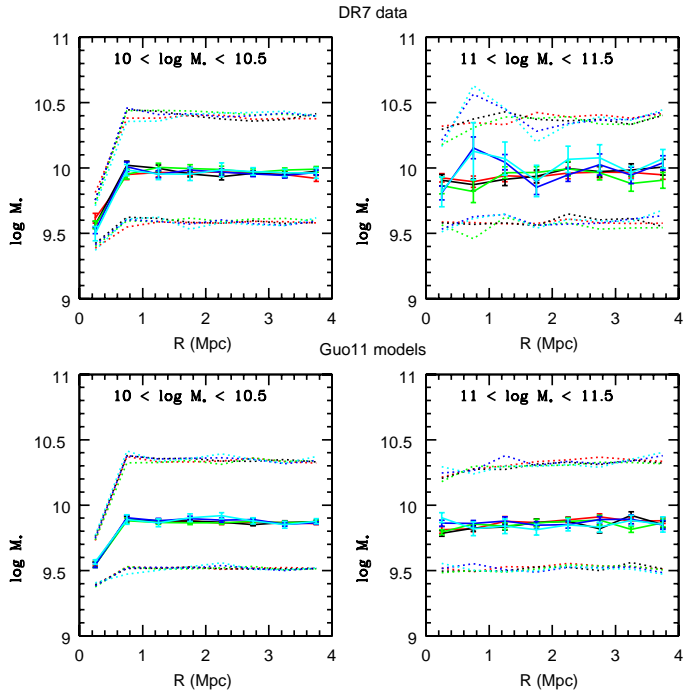}
\caption{ The median stellar masses of  
neighbouring galaxies (solid lines), as well as the
25th and 75th percentiles of the stellar mass distributions of these objects (dotted lines),
are plotted as a function of projected distance from
central galaxies in the model. Results are shown for central galaxies in two stellar mass ranges. 
The central
galaxies have been ordered by cold gas mass fraction and colour-coded
as in previous figures. Results from the SDSS DR7 are shown
in the top two panels, while results from the Guo et al. (2011) models are shown in the
bottom panels. 
\label{sample}}
\end{figure}

Finally, we would like to note that it would
be  useful to search for direct evidence of pre-heating of the gas around
red, low mass galaxies  using quasar absorption lines as probes of the
temperature of the surrounding IGM.  It will also be  important to understand the physical
processes responsible for the heating. The strong correlation between gas content in
low mass central galaxies and the timescale over which the neighbouring galaxies
are building their {\em central} stellar populations offers a tantalizing hint that
feedback processes associated with galaxy bulge and black hole formation may play a role.  
This will be the subject of future work.

\section*{Acknowledgments}
Thank you to the Aspen Center for Physics and the NSF Grant No. 1066293 for hospitality,
support and peace-of-mind during the conception and writing of this paper.


\end{document}